# Impact of surfactant polydispersity on the phase and flow behaviour in water: the case of Sodium Lauryl Ether Sulfate

Rosalia Ferraro[a,b], Maria Michela Salvatore[c], Rodolfo Esposito[c], Sergio Murgia[d,e], Sergio Caserta[a,b,*], Gerardino D'Errico[c,d, †], Stefano Guido[a,b, †]

[a] DICMaPI, Università di Napoli Federico II, P.le V.Tecchio 80, 80125 Napoli, Italy

[b] CEINGE Advanced Biotechnologies Franco Salvatore, Via Gaetano Salvatore, 486, 80131 Napoli, Italy

[c] DSC, Università degli Studi di Napoli Federico II, Via Cintia, 21, 80126 Napoli, Italy

[d] CSGI, Consorzio Interuniversitario per lo Sviluppo dei Sistemi a Grande Interfase, Via della Lastruccia 3, I-50019 Sesto Fiorentino, Florence, Italy

[e] Department of Life and Environmental Sciences, University of Cagliari, Cittadella Universitaria Monserrato, S.P. 8 Km 0.700, 09042 Monserrato (CA), Italy

† GDE and SG contributed equally to this work.

* Correspondence should be addressed to: Sergio Caserta, DICMaPI, Università di Napoli Federico II, P.le V.Tecchio 80, 80125 Napoli, Italy,
sergio.caserta@unina.it

This research did not receive any specific grant from funding agencies in the public, commercial, or not-for-profit sectors.

# Graphical abstract

To investigate the effects of Sodium Lauryl Ether Sulfate (SLES) polydispersity on its phase behaviour, we performed Time-Lapse dissolution experiments within microchannel geometries. These studies allowed us to observe the dynamic interactions between a surfactant sample and water, examining the surfactant's phase transitions in real-time. Our observations unveiled a coexistence region within the phase diagram, characterized by distinct microstructures. Detailed analyses identified these as hexagonal, cubic, and lamellar molecular formations. Remarkably, this microstructural coexistence – permitted by the local segregation of surfactant fractions varying in hydrophilicity – gives rise to unique structural and flow properties in the aqueous surfactant mixture.

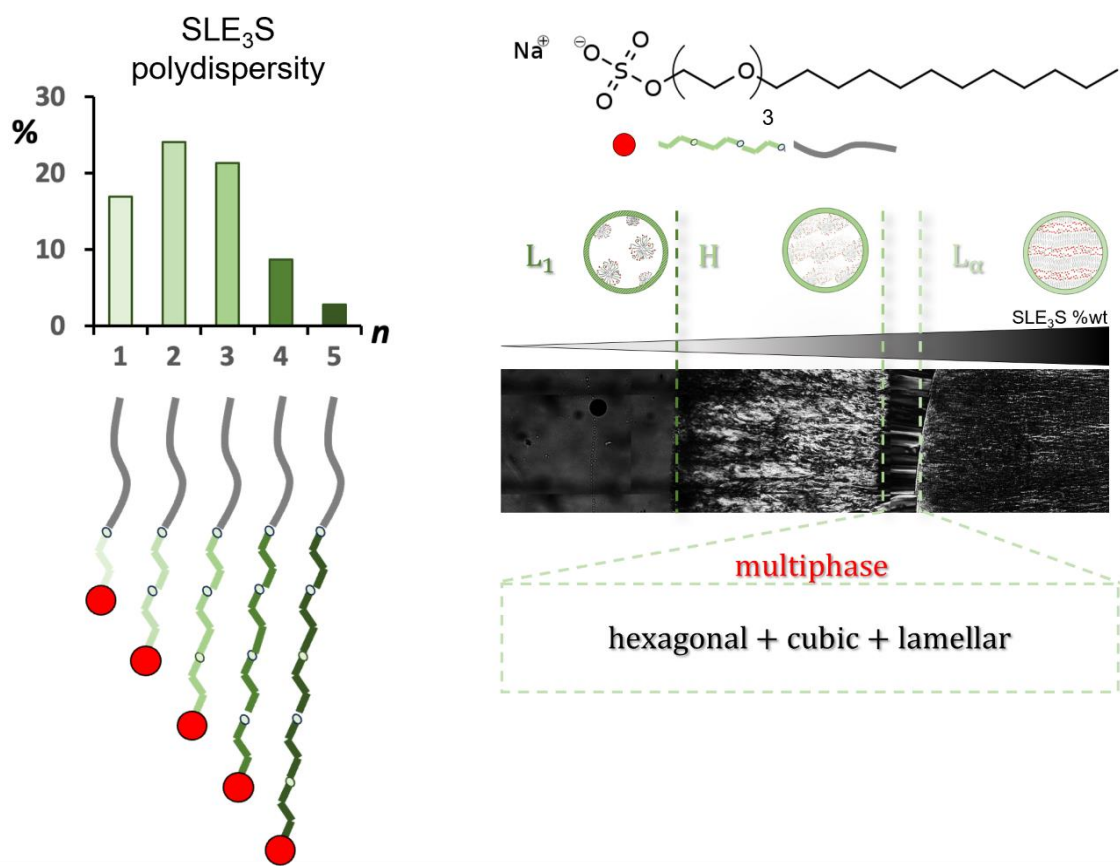

# Abstract

This study delves into the impact of molecular polydispersity on the phase behaviour of Sodium Lauryl Ether Sulfate (SLES) surfactant, aiming to deepen understanding of its implications for fundamental science and industrial applications. $SLE_3S$ is utilized as a model compound: a comprehensive characterization of molecular polydispersity is conducted using Gas Chromatography–Mass Spectrometry and Nuclear Magnetic Resonance spectroscopy, juxtaposing the findings with those for $SLE_1S$.

Our comprehensive investigative approach entails: (i) employing Time-Lapse dissolution experiments in microchannel geometries to observe the dissolution and phase transitions; (ii) utilizing polarized light microscopy, confocal microscopy, and Small Angle X-ray Scattering for microstructure identification and concentration assessments; (iii) conducting rheological evaluations at various concentrations and temperatures to determine their effects on the surfactant properties.

The findings reveal that $SLE_3S$, being more polydisperse, demonstrates complex phase behavior not observed in the less polydisperse $SLE_1S$. Notably, $SLE_3S$ exhibits a unique concentration domain, corresponding to a concentration of 60%wt, where hexagonal (H), cubic, and lamellar ($L_\alpha$) phases coexist, resulting in highly viscoelastic heterogeneous mixtures. This behavior is attributed to the local segregation of surfactant components with varying polarity, underscoring the crucial role of molecular polydispersity in the phase behavior of SLES surfactants.

# 1. Introduction

Surfactants play an indispensable role in shaping various aspects of our daily lives [1]. These versatile compounds are widely present in household and personal care products [2, 3]. Moreover, surfactants find extensive applications as stabilizers in paints and plastics, as well as effective corrosion inhibitors [4-6]. Beyond these conventional uses, their unique properties position them as promising candidates for revolutionizing biological processes, such as drug extraction and delivery [7-16], thanks to the formation of mesophases when mixed with water. Understanding the phase and flow behavior of aqueous surfactant systems and how it is affected by concentration and temperature has profound implications, not only for shaping the properties of the final product [17-19], but also for crucial processing steps like dissolution in water [20]. Scientific and technological research in the field is generally based on the combination of several experimental techniques and on the close scrutiny of a large number of samples.

In the great majority of studies, monodisperse surfactant samples are considered, being polydispersity regarded as a useless complication hampering unambiguous result interpretation. Contrarily, industrial practices often employ commercial surfactants with considerable molecular polydispersity – a factor whose effects are typically overlooked. This work demonstrates that molecular polydispersity is pivotal in dictating the phase and flow behaviors of surfactants, which could markedly diverge from those of their monodisperse counterparts.

As a relevant case study, we focus on Sodium Lauryl Ether Sulfate (SLES), which is one of the most widely produced and employed surfactants worldwide, being considered a rather un-expensive and eco-friendly anionic surfactant. The SLES phase behavior has been the subject of various investigations [21, 22], but the polydispersity role has not received the due attention. Commercial SLES samples present a complex polydispersity arising from at least three factors: i) the number of ethoxylic units, ii) the length of the alkyl tail and iii) the sulfonation degree. Prior research about the relation between molecular polydispersity and phase behavior is limited to the behavior of polydisperse alkyl ethoxylates, thus only focusing on the number of ethoxylic units. Pseudo-binary phase diagrams qualitatively resembling those of monodisperse surfactants are reported [23]. However, more quantitative comparisons have uncovered distinct characteristics: firstly, polydisperse surfactants tend to be more hydrophobic than monodisperse surfactants with the same (average) number of ethoxylate units, underscoring the significant role played by unreacted alcohol and ethoxylated alcohols with low levels of ethylene oxide in governing surfactant self-assembly [24, 25]. Secondly, in the pseudo-binary phase diagrams of polydisperse surfactant-water mixtures, three-phase coexistence regions can be observed, while true binary phase diagrams of monodisperse surfactant-water mixtures allow three-phase coexistence only at a single point [23]. It is also important to notice that surfactant polydispersity has been recognized as a significant factor influencing phase behavior in pseudo-ternary mixtures involving water and a hydrophobic solvent (an oil) capable of forming microemulsions. In these systems, polydisperse alkyl ethoxylates segregate, with the more lipophilic surfactant fraction solubilized by the oil phase, while the remaining one positions at the water/oil interface. Consequently, the surfactant appears more hydrophilic on average than expected due to its negligible solubility in water [26-30]. This phenomenon has also been quantitatively analyzed through the intentional mixing of monodisperse surfactant samples at known ratios [31-34]. Finally, surfactant polydispersity has been found to play a key role in the formation of a sponge phase in mixtures comprising a commercial polyethoxylated non-ionic surfactant, a cosurfactant (ethylhexyl glyceryl ether), and water [35].

Given the compelling evidence of the influence of molecular polydispersity in non-ionic surfactant phase behavior, this study represents the first in-depth investigation of molecular polydispersity's effects in SLES-water systems. The surfactant sample used in this study is described by the average chemical formula $C_{12}H_{25}(OCH_2CH_2)_3OSO_3Na$, indicating an average ethoxylation degree of approximately three ($SLE_3S$). As a preliminary step in this study, we conducted a comprehensive quantitative characterization of the sample's polydispersity using Gas Chromatography–Mass Spectrometry (GC-MS) and Nuclear Magnetic Resonance (NMR) spectroscopy and compared the results with those obtained for $SLE_1S$. To quantify the impact of polydispersity on the phase diagram, we conducted Time-Lapse dissolution experiments in a microchannel geometry to compare the dissolution behavior of these samples when exposed to water and, subsequently, to examine the resulting phase transitions.

As a result, we observed that $SLE_3S$ exhibited one additional microstructure that was absent in $SLE_1S$. To identify the specific concentration and type of microstructure, we employed a combination of experimental techniques, including polarized light visual inspections, confocal microscopy, and Small Angle X-ray Scattering (SAXS). Furthermore, to quantify differences in terms of viscoelastic properties, we conducted rheological characterizations at varying concentrations of $SLE_3S$ as a function of temperature. The outcomes were interpreted in light of the surfactant molecular composition, thereby elucidating the effects of polydispersity.

## 2. Materials ad methods

### 2.1. Comparative Spectroscopic and Spectrometric Profiling: SLE$_3$S and SLE$_1$S

The SLE$_3$S paste utilized in this research was supplied by Procter and Gamble, containing a surfactant concentration of 72% by weight, verified through vacuum drying. SLE$_3$S, a sodium salt formed from sulfated ethoxylated alcohols, is declared by the supplier to have an average of three ethoxylic groups [7].

The chemical composition of the SLE$_3$S paste was spectroscopically characterized by $^1$H NMR analyses. $^1$H NMR spectra were recorded on a Bruker AMX instrument (Rheinstetten, Germany) at 400 MHz in D$_2$O (99%, Sigma, Milan, Italy) at 25 °C. The same solvent was used as internal standard. Complementing this, SLE$_3$S pastes were analyzed using GC-MS [36]. For that, SLE$_3$S samples were dried under a stream of nitrogen and accurate weight amounts were dissolved in methanol. 2 μL of each sample solution were injected and analyzed by an Agilent 6850 GC (Milan, Italy), equipped with an HP-5MS capillary column (5% phenyl methyl polysiloxane stationary phase), coupled to an Agilent 5973 Inert MS detector operated in the full scan mode (*m/z* 35–550) at a frequency of 3.9 Hz and with the EI ion source and quadrupole mass filter temperatures kept, respectively, at 200 and 250 °C. Helium was used as carrier gas at a flow rate of 1 mL·min$^{-1}$. The injector temperature was 280 °C and the temperature ramp raised the column temperature from 70 to 280 °C: 70 °C for 1 min; 10 °C·min$^{-1}$ until reaching 170 °C; and 30 °C·min$^{-1}$ until reaching 280 °C. Then, it was held at 280 °C for 5 min. The solvent delay was 4 min. The identification was performed by matching their EI mass spectra at 70 eV with those stored in the NIST20 mass spectral library [37]. Furthermore, the identification was supported by the Kovats retention index (RI) calculated for each metabolite by the Kovats equation using the standard *n*-alkane mixture in the range C7-C40 (Sigma) analyzed under the same conditions.

For comparative analysis, a SLE$_1$S paste, also provided by Procter and Gamble, underwent identical spectroscopic examination, setting the stage for subsequent comparative evaluations.

### 2.2. Integrative Methodologies for Comprehensive Phase Diagram of SLE$_3$S-water system

In accordance with the typical behavior of surfactant systems, SLE$_3$S displays a series of phase transitions influenced by variations in concentration and temperature. At room temperature (25 °C) previous research [20] have identified several phases within specific concentration ranges: 0.024-28%wt - micellar, 31-56%wt - hexagonal, ~58-62%wt - cubic, and 62-72%wt - lamellar. Despite initial phase identifications, a detailed temperature-concentration phase diagram for this surfactant remains undocumented, a significant omission given the compound's scientific and industrial relevance. This study aims to contribute to understand the phase behavior by integrating a range of experimental methodologies: i) Time-Lapse microscopy in a microfluidic channel; ii) confocal and polarized light microscopy; iii) visual inspection; iv) SAXS measurement and v) rheological characterization.

*2.2.1. Sample preparation*

Mixtures at different concentrations of SLE$_3$S were prepared adding the required amount of bi-distilled water to the raw surfactant paste. Each sample was thoroughly mixed with a spatula during dilution and then subjected to centrifugation (at 4000 rpm for 40 minutes) multiple times to remove any trapped gases. The samples were stored at room temperature overnight to guarantee they reached equilibrium. For confocal imaging (as described in **section 2.2.3**), samples were stained by adding Rhodamine B (Sigma, Milan) as a fluorescent marker at a low concentration ($10^{-3}$ mg/mL) to minimize possible interference with the phase behavior.

*2.2.2. Time-Lapse dissolution in a microchannel geometry*

Polarized optical microscopy (POM) was employed to observe the dynamic evolution of both surfactant solutions during dissolution in water in a microchannel geometry configuration, and, subsequently, to examine the resulting phase transitions. A rectangular microchannel of height 200 µm and width 5 mm, obtained by modifying a commercial bottomless chamber (Ibidi sticky Slide I Luer) [38], was used. The so obtained device consisted of two openings: one used to introduce the surfactant and another with a small pool created by using an O-ring, allowing contact with bi-distilled water. The dissolution experiments were conducted, both for SLE$_1$S and SLE$_3$S, at different temperatures: 30, 40, 50 and 60 °C. Images were acquired at regular intervals (*i.e.*, 10 minutes) until full dissolution.

SLE$_1$S and SLE$_3$S phases evolution was followed by automated Time-Lapse Microscopy (TLM) based on an inverted microscope (Zeiss Axiovert 200, Carl Zeiss, Jena Germany). Several independent fields of view were acquired by a high-resolution high-sensitivity monochromatic CCD video camera (Hamamatsu Orca AG, Japan) using a 10x objective and two crossed polarizers to visualize the internal microstructure. The workstation was also equipped with a motorized stage and focus system (Marzhauser) controlled with a home-made LabVIEW code for automated mosaic scanning (*e.g.*, 5x4 images) of large samples.

*2.2.3. Confocal and polarized light microscopy*

Images were collected at room temperature using a Laser Scanning Confocal Microscope (LSM) 5 Pascal (Carl Zeiss Advanced Imaging Microscopy, Jena, Germany). A small amount of surfactant at different concentrations was squeezed between a glass microscope slide and a coverslip. To control the sample thickness, a double-sided adhesive tape was placed as a spacer between the two glass surfaces, resulting in a thickness of 130 µm. Observations were made between crossed polarizers for light microscopy, and without polarizers for confocal microscopy.

*2.2.4. Visual inspection*

Visual inspection was performed at different temperatures (from room temperature to 60 °C). Samples at different concentrations were loaded in small cylindrical glass bottles (volume ~ 4 mL) and centrifuged to degas. These bottles were placed in a rectangular glass pool previously filled with silicon oil (200 cSt). The latter, with a refractive index close to that of glass (1.40 and 1.51, respectively), was used to minimize curvature effects arising from the cylindrical shape of the bottles. Within the same pool, two bottles containing bi-distilled water and air were also included as control samples. Two flat sheet polarizers were sticked to two opposite walls of the pool and crossed with each other. Different orientation angles of the front polarizer with respect to the vertical direction are considered (0, 45, 90 and 135°), while keeping the back polarizer (analyser) always crossed with respect to the front one (+90°). Typical images are reported in **Figure S1** in the **Supplementary Materials**. The entire pool, containing the samples in the silicon oil bath, was placed on an electric hotplate to control the temperature, that was measured with a thermocouple, and was imaged using a Canon EOS 60D camera (Amstelveen, The Netherlands) with an objective Canon EF-S 60mm f/2.8 Macro USM placed on a tripod and carefully aligned to be orthogonal to the front wall of the pool.

*2.2.5. SAXS measurements*

X-ray scattering is the most commonly used method for phase identification, assigning each peak in the sample pattern to a phase. Diffraction experiments were performed using a SAXS setup, in order to obtain verification of phase assignments from POM. X-ray scattering patterns were recorded using a S3-Micro SWAXS camera system (HECUS, Graz, Austria) employing Cu Kα radiation ($\lambda = 1.542$ Å). The S3 Micro system comprises a GeniX X-ray generator, which operated at 50 kV and 1 mA. The scattered X-rays were detected by a 1D-PSD-50M system containing 1024 channels of width 54.0 µm. Detector covered the s-range of interest from about 0.02 to 0.4 Å$^{-1}$ ($q = (4\pi/\lambda)\sin(\theta)$, where $2\theta$ is the scattering angle). Silver behenate, $CH_3$–$(CH_2)_{20}$-COOAg, was used as a standard to calibrate the angular scale of the intensity, $I(q)$. A stainless-steel sample holder with thin polymeric sheet beam window (Kapton X-ray film roll TF-475, FluXana GmbH & Co. KG, Bedburg-Hau, Germany) was used. Intensity was measured not only at room temperature, but also at 40, 50 and 60 °C. A Peltier element is used to control temperature (±0.1 °C).

*2.2.6. Rheology setup*

Rheological measurements were performed for SLE$_3$S solutions at different concentration (20, 35, 50, 60 and 72%wt) and at different temperatures (30, 40, 50 and 60 °C) by using a stress-controlled rheometer (Anton Paar Physica MCR 301 Instruments, Graz, Austria) equipped with a cone-plate measuring geometry (CP25-1/S-SN72510). Temperature control was achieved through a Peltier cooler/heater connected to an external circulating water bath (Lauda, Milan, Italy). The measurements were run according to a protocol previously applied to surfactant systems [39]. Prior to the measurements, the samples underwent a pre-shear (1 min at 200 1/s) to cancel out possible loading effects. Flow curves were obtained by varying shear rate, $\dot{\gamma}$, in the range 0.01–200 1/s under steady shear flow. Oscillatory flow tests were conducted within the linear viscoelastic regime. To confirm linear viscoelasticity, strain sweep tests were performed at 10 rad/s, with the amplitude ranging from 0.01 to 1. Frequency sweep tests are performed at strain 0.1 and varying frequency in the range 0.1-100 rad/s.

## 3. Results and Discussion

### 3.1. NMR and GC-MS Insights into SLE$_1$S and SLE$_3$S

NMR spectroscopy was initially utilized for the structural elucidation of surfactants. The $^1$H NMR spectra of SLE$_1$S and SLE$_3$S exhibited the characteristic resonances associated with sodium lauryl sulfate (SLS) and SLES (refer to **Figure S2** and **Table S1**). Through signal integration within these spectra, we quantified the composition of the samples (detailed methodology is provided in the **Supplementary Material**). Our analysis confirmed that the average ethoxylation degree of SLE$_1$S and SLE$_3$S pastes are approximately 0.8 and 2.4, respectively, in fair agreement with the data furnished by the provider. The fraction of SLS (*i.e.*, non-ethoxylated molecules) in SLE$_1$S and SLE$_3$S pastes are approximately 0.6 and 0.2, respectively. Finally, the fraction of non-sulfonated molecules (*i.e.*, fatty alcohols and alkyl ethoxylates) in SLE$_1$S and SLE$_3$S pastes are approximately 0.1 and 0.3 in SLE$_1$S and SLE$_3$S, respectively.

To characterize surfactants with respect to the alkyl and ethylene oxide chain distribution, GC-MS was employed for the analyses of SLE$_1$S and SLE$_3$S pastes. GC-MS results of methanol solutions of SLE$_1$S and SLE$_3$S pastes showed the presence of thermal degradation products of both SLS and SLES (refer to **Table S2**). In fact, it is well-known that sulfonate compounds are quantitatively converted in their de-sulfonated products at the high temperature of the GC injector [36]. Hence, we employed the peak area percent of the resulting ethoxylated alcohols for the following determinations. As expected, our data showed that SLE$_1$S and SLE$_3$S pastes were constituted by alkyl chains with 12-16 carbon atoms and the most abundant was the dodecyl chain in both samples under examination (**Figure 1A**). The fraction of SLS in SLE$_1$S and SLE$_3$S pastes was 0.51 and 0.17 respectively, thus confirming the results obtained by NMR. The number of ethoxylic units of SLE$_1$S and SLE$_3$S is quite polydisperse (**Figure 1B**): while the average ethoxylation degree is 0.69 and 2.0, respectively, it has to be stressed the presence of up to penta-ethoxylated species in the SLE$_3$S sample.

Summarizing, the quantitative analysis of the sample composition and polydispersity shows that the SLES samples used for this study contained a large spectrum of compounds, including significant fractions of non-ethoxylated alkyl sulfate. Particularly, the SLE$_3$S sample also contained a large amount of non-sulfonated alkylethoxylates.

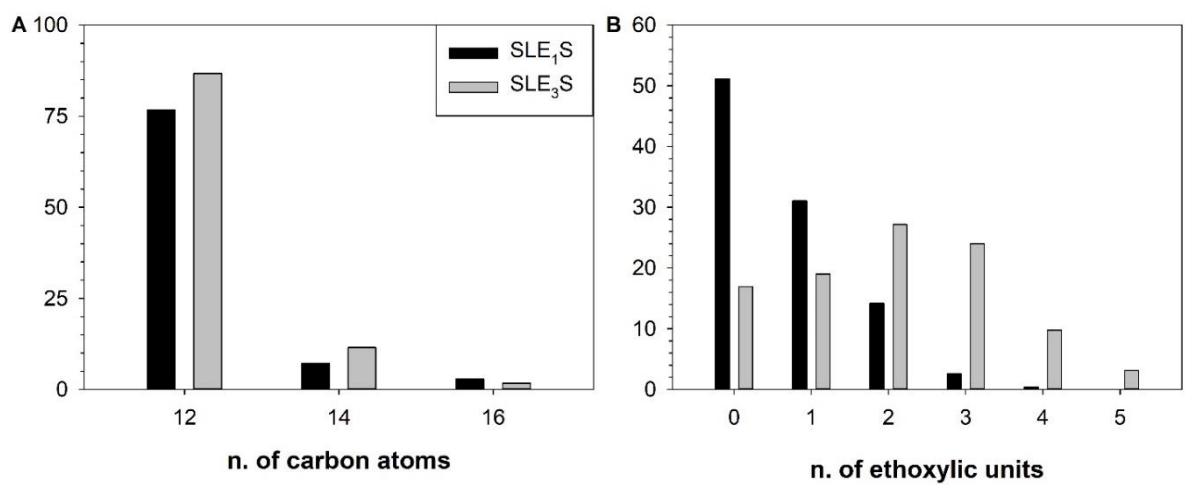

*Figure 1. GC-MS characterization of SLE$_1$S and SLE$_3$S surfactants. (**A**) Distribution of the number of carbon atoms of the surfactant tails and (**B**) the number of ethoxylic units in the hydrophilic headgroups for SLE$_1$S (black) and SLE$_3$S (grey).*

## 3.2. Microstructural Dynamics in SLES Systems: Insights from Time-Lapse Dissolution Studies

**Figure 2A** illustrates the dissolution of SLE$_1$S at 72%wt and 30 °C. Over 30 minutes, three distinct strips emerged: a dark strip on the left indicating the L$_1$ phase at low surfactant concentration, a brighter central strip representing the H phase, and a rightmost strip with Maltese crosses showing the concentrated L$_\alpha$ phase. Yellow dotted lines in the figure demarcate the phase boundaries. Conducting the same experiment under identical concentration and temperature conditions with SLE$_3$S, as depicted in **Figure 2B**, revealed four distinct strips, indicating an additional phase absent in SLE$_1$S. From left to right, the sequence begins with the dark L$_1$, followed by the more birefringent H and then a phase likely corresponding to the isotropic cubic phase, distinguished by alternating dark and bright horizontal bands. This suggests a coexistence of birefringent structures alongside the cubic phase. The final strip, also in this case, is a L$_\alpha$, characterized by typical Maltese cross patterns, as reported in the insets.

Extending our analysis to examine temperature effects on these microstructures, we conducted the SLE$_3$S dissolution experiment at higher temperature. The POM images in **Figure 2C** show that while the L$_1$ and L$_\alpha$ phases remain largely unchanged at 60°C, the multi – structured phase structure darkens, hinting at a potential phase transition. Notably, the bright bands previously seen in the multi – structured phase are no longer visible, suggesting a microstructural transformation. Additionally, darker, ellipsoidal regions can be indeed noticed in the H phase, likely due to nucleation of a isotropic phase. This comprehensive analysis underscores the significant role of both polydispersity and temperature in influencing the microstructural behavior of these surfactant systems.

To transform our qualitative observations into quantitative data, we analyzed the light intensity profile (refer to the bottom of **Figure 2**). This was achieved by averaging the grey level ($I$) along the y-direction at each x position and normalizing it against the intensity of pure water ($I_w$). The x position is measured with respect to the $L_1$-H interface, which is dynamically moving from the water reservoir to the bulk of the microchannel during the dissolution experiment [22, 38, 40]. For both $SLE_1S$ and $SLE_3S$ at 30 °C, the normalized light intensity profile exhibits a noisy pattern, indicative of a polydomain microstructure. Each microstructure is distinguished by a unique average value, allowing for a clear correlation between the observed images and the corresponding trends. The phase boundaries within these images are marked by white dotted lines for clarity. At higher temperatures, however, the profile smoothens, implying a reduction in local ordering, particularly within the H phase.

The identification of the cubic phase, previously noted in literature but not structurally analyzed, has propelled us to delve deeper into this aspect of the $SLE_3S$ phase diagram. The ensuing sections will elaborate on this analysis, providing a more comprehensive understanding of the phase characteristics and behaviors in these surfactant systems.

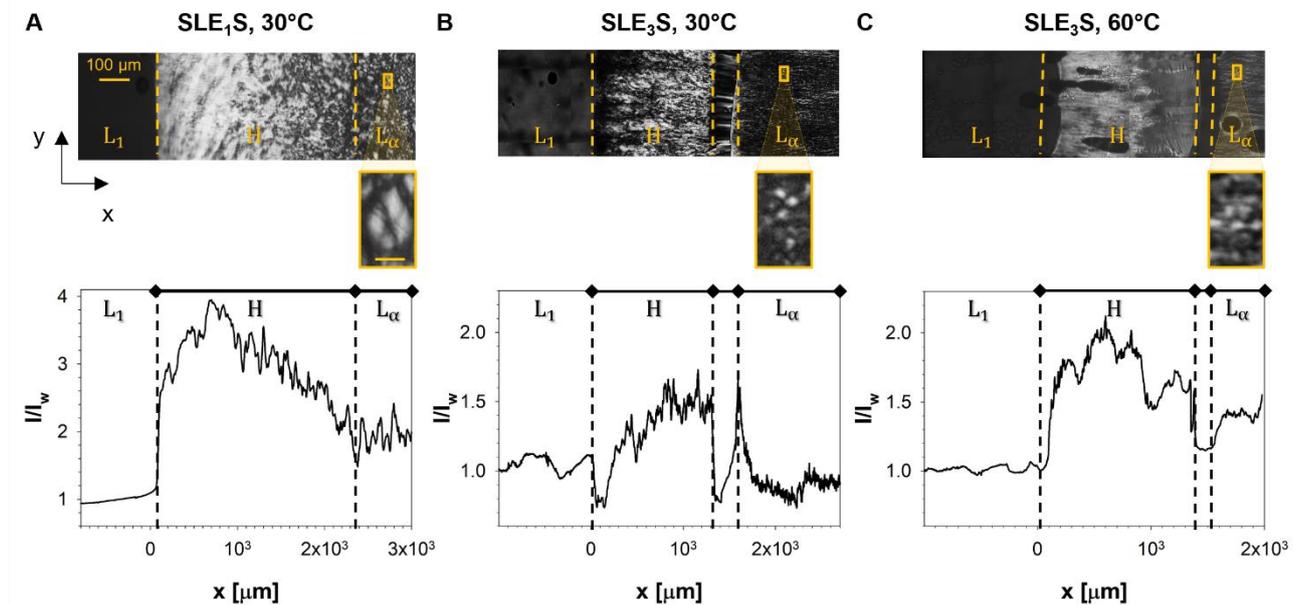

*Figure 2. Phase behavior of surfactant systems.* *Polarized Optical Microscopy (POM) images of $SLE_1S$ at 30°C (**A**), $SLE_3S$ at 30°C (**B**), and $SLE_3S$ at 60°C (**C**). The mean grey levels (I), normalized with respect to the intensity in pure water ($I_w$), were plotted against the position on the x-axis (µm). Phase boundaries are indicated by yellow dashed lines. Each phase, including micellar ($L_1$), hexagonal (H), multi – structured, and lamellar ($L_α$), is represented on the graph in distinct sections demarcated by dashed lines and diamonds. Insets provide a magnified view of the Maltese cross, indicative of the lamellar phase. The scale reported in the figure is 50 µm.*

## 3.3. Microscopic Characterization of Phase Transitions in SLE$_3$S

In **Figure 3**, the top row displays representative polarized optical microscopy images of SLE$_3$S solutions at concentrations at 25 °C: from left to right, the columns correspond to 55%, 60%, 61%, and 70%wt concentrations. The H phase, corresponding to a concentration of 55%wt, exhibits a striated bright irregular nongeometrical pattern, due to its anisotropic nature. On the contrary, samples at 60 and 61%wt appear largely black, being an isotropic cubic phase, except for isolated birefringent domains [20, 22], suggesting coexistence of different phases. The sample at 70%wt shows the birefringent texture typical of the L$_\alpha$ phase with Maltese crosses, highlighted by yellow arrows in **Figure 3**.

We further explored solution heterogeneity using Rhodamine B, as depicted in the second row of **Figure 3**. A clear correlation between structures observed with POM and with confocal microscopy is recognizable, such as the striated pattern and the Maltese crosses observed for the H and L$_a$ phases, respectively. The solutions at 60 and 61%wt reveal a combination of brighter and darker domains of irregular shape; these images further support the coexistence of multiple microstructures alongside the cubic phase. In fact, the varying brightness across different domains is attributed to the differential affinity of Rhodamine B for each microstructure, providing additional insights into the complex nature of these phase transitions.

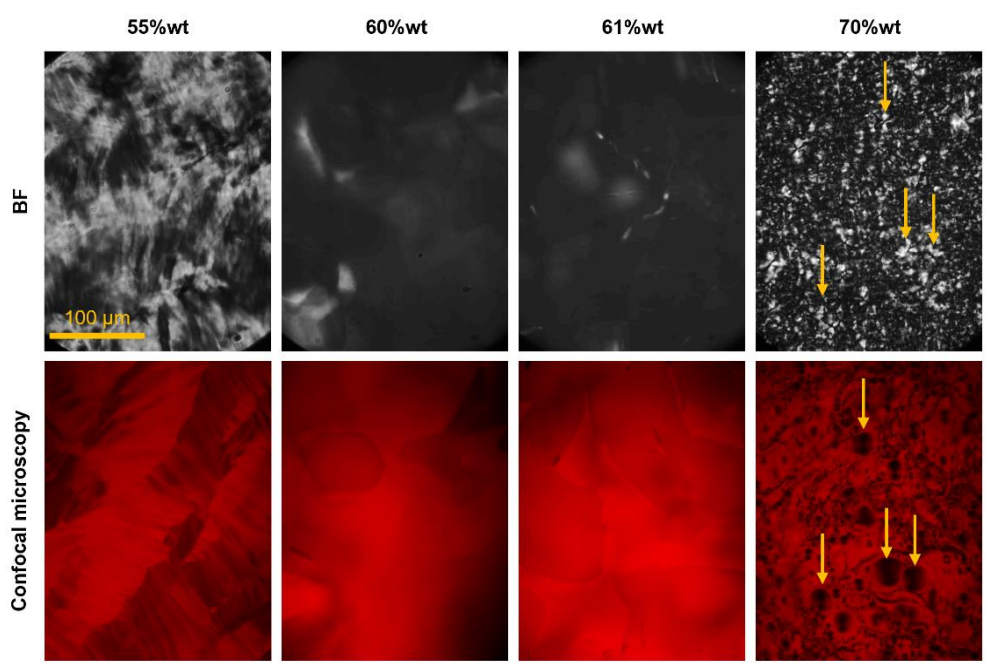

*Figure 3. Microstructural analysis of SLE$_3$S solutions at various weight percentages.* *The top row (BF = Bright Field) displays representative polarized optical microscopy images of SLE$_3$S solutions at concentrations of 55%, 60%, 61%, and 70%wt under crossed polarizers, revealing the structures characteristic of hexagonal (H), multi – structured, and lamellar (L$_\alpha$) phases. The bottom row showcases confocal microscopy images for corresponding concentrations of SLE$_3$S. All images were captured at 25°C. The yellow arrows in the 70%wt images highlight the presence of Maltese crosses, a feature typical of the lamellar phase.*

## 3.4. Macroscopic Characterization of Phase Transitions in SLE$_3$S

Shifting from microscopic to macroscopic analysis, typical macro-photography images are presented in **Figure 4A**. Each row of the table is relative to a different sample temperature (25, 45 and 60 °C) while in each column a sample with a different concentration (35, 50, 60 and 72%wt) is shown, alongside air and water (0%wt of SLE$_3$S) samples referred to as a control.

At room temperature, all solutions observed show a bright appearance against the dark background. In particular, solutions at 35 and 50%wt, corresponding to the H phase [20], and at 60%wt, denoting the cubic phase [20], are notably brighter than the 72%wt, L$_\alpha$ phase [20]. As expected, both air and water samples appear black. The brightness observed in hexagonal and lamellar surfactant mixtures can be attributed to the presence of anisotropic, birefringent structures. Interestingly, the brightness remains largely unchanged irrespective of the orientation of the crossed polarizers with respect to the samples (see **Supplementary Materials**), suggesting the presence of a polydomain structure, characterized by liquid crystalline regions with different orientation within the sample.

More puzzling is the bright appearance of the 60%wt solution since a cubic phase should be isotropic and therefore appear dark under crossed polarizers. A possible explanation is that the brightness comes from the small birefringent domains observed in POM micrographs, corresponding to the bright bands visible in the multi – structured domain of TLM images.

Upon increasing the temperature from 25 to 45 °C, no remarkable changes can be observed in all the samples; conversely, by rising temperature up to 60 °C, the cubic phase solution, 60%wt, shows a drastic reduction in birefringence, compared to the other phases. On the contrary, hexagonal and lamellar structures are unaffected by temperature increments from 25 to 60 °C. This observation hints at a potential phase transition occurring in the 60%wt SLE$_3$S solution below 60 °C, underscoring the temperature-sensitive nature of the surfactant phases.

**Figure 4B** quantitatively captures the change in light intensity profiles, *I*, across different temperatures for each examined concentration. As we escalate the temperature from 25 °C to 45 °C, the birefringence of the samples remains largely unaffected. However, a notable change is observed at 60 °C for the 60%wt concentration, where the multi – structured phase exhibits a significant decrease in birefringence, plunging from an intensity value of 141 to just 54. The inset of **Figure 4B** presents the intensity profile for the 60%wt sample at 45 °C, plotted against the bottle height, to accentuate the gradual phase transition induced by heating.

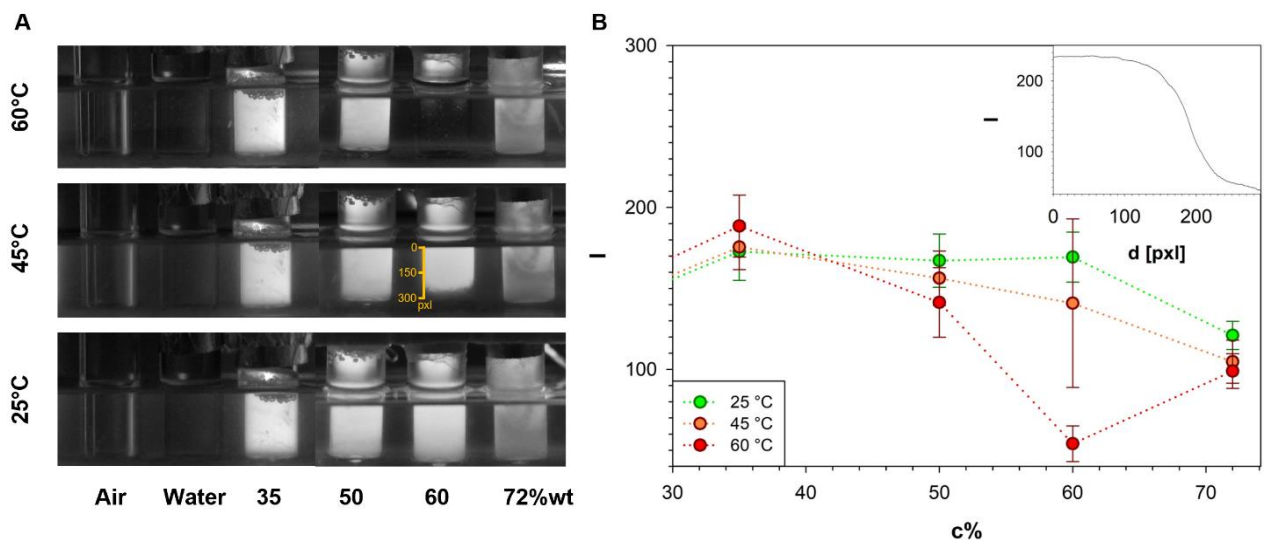

*Figure 4. Temperature and concentration-dependent behavior of SLE$_3$S. A) Visual observations of SLE$_3$S across a range of temperatures (25°C, 45°C, and 60°C) and concentrations (35%, 50%, 60%, and 72%wt). The samples are arrayed in rows by acquisition temperature and in columns by SLE$_3$S concentration, with air and water serving as controls for comparison. B) Optical intensity as a function of SLE$_3$S concentration at the different temperatures (25°C – green, 40°C – orange and 60°C – red), with the inset showing the intensity profile of SLE$_3$S at 60%wt and at 45°C against pixel distance (d [pxl]), from the top to the bottom of bottle.*

### 3.5. SAXS Analysis of SLE$_3$S

SAXS analysis was conducted to elucidate the multi-structured phase characterizing SLE$_3$S, with the results displayed in **Figure 5A**. This figure presents the SAXS profiles for various SLE$_3$S concentrations (50, 59, 60, 61, and 72%wt) at room temperature.

For 50%wt SLE$_3$S solutions, two Bragg peaks are detected. The second reflection at q=0.206 Å$^{-1}$ identifies a hexagonal arrangement of the liquid crystal, being in a proportion of $4^{0.5}:3^{0.5}$ with the scattering vector of the first reflection (q=0.119 Å$^{-1}$), highlighted in orange in the figure. Conversely, for 72%wt SLE$_3$S solution, a distinct reflection occurs at q=0.286 Å$^{-1}$, identifying a lamellar phase, with a ratio of 2:1 to the first peak (q=0.143 Å$^{-1}$). Both peaks are highlighted in red in the plot.

In samples identified by POM as cubic phases, intricate reflection patterns suggest the presence of multiple liquid crystalline phases. At 59%wt, the Bragg peak at q=0.125 Å$^{-1}$ and a second one at q=0.217 Å$^{-1}$, related in a proportion of $4^{0.5}:3^{0.5}$, identify the hexagonal phase. At the same time the reflection at q=0.132 Å$^{-1}$, related to the shoulder at q=0.264 Å$^{-1}$ in a proportion of 2, corresponds to the L$_\alpha$ phase. Along with these peaks, other (relatively weaker) peaks are detected, which are ascribable to the cubic supramolecular arrangement. Given that numerous surfactants are known to form a bi-continuous V phase situated between the stability domains of the H and L$_\alpha$ phases, it was hypothesized that the results could be indicative of bi-continuous cubic structures. Nevertheless, efforts to correlate the observed SAXS reflections with the space groups commonly associated with such bi-continuous structures did not yield conclusive matches. Instead, the presence of a peak at

q=0.107 Å$^{-1}$, along with reflections at q=0.205 and 0.247 Å$^{-1}$ (in green), proportional to each other by factors $11^{0.5}:3^{0.5}$ and $16^{0.5}:3^{0.5}$, suggests the formation of a cubic phase with Fd3m symmetry. This symmetry is characteristic of a discontinuous arrangement of micellar aggregates, a finding that is particularly surprising given that such liquid crystalline phases are generally observed at lower surfactant concentrations, in the region between the L$_1$ and H phases.

Upon a marginal increase in concentration from 59 to 60%wt, the SAXS profile maintains its complexity. Notably, within the cubic phase, a third reflection at q=0.173 Å$^{-1}$, related to the first peak in a proportion of $8^{0.5}:3^{0.5}$, which lends additional support to the presence of Fd3m symmetry. Nonetheless, the exact structural configuration of the cubic phase in SLE$_3$S remains to be definitively determined through further study.

Across the concentrations examined, the phase diagram for SLE$_3$S-water mixtures consistently shows the coexistence of H, cubic, and Lα phases within a specific concentration window. This coexistence seems to challenge the traditional interpretations of the Gibbs phase rule. As we continue to increase the surfactant concentration beyond this window, the hexagonal phase is no longer observed, leaving only the Lα and cubic phases detectable in the 61%wt sample.

The temperature dependent SAXS profiles for the 60%wt SLE$_3$S sample are presented in **Figure 5B**. Specifically, at 40 °C, the H phase fully dissolves, while the Fd3m phase gradually diminishes and eventually vanishes at 50°C. Conversely, the L$_α$ phase persists until 60°C when it completely disappears. These observations affirm that the SLE$_3$S-water system functions as a multi-component system. As detailed in section 3.1, the SLE$_3$S sample contains molecules with largely variable degree of ethoxylation, which means variable hydrophilicity and packing attitude. Moreover, there is a significant presence of non-sulfonated alkylethoxylates. We suggest that the polydispersity of the surfactant plays a pivotal role in dictating its phase behaviour, which could be related to the segregation of surfactant fractions with different molecular features: as the concentration is raised above the stability domain of the H phase, surfactant molecules constituted by long alkyl and short ethoxylic chains start to form L$_α$ structures, while the molecules constituted by short alkyl and long ethoxylic chains remain in the hexagonal arrangement. The cubic phase could be formed by molecules with intermediate hydrophilicity (if it will be confirmed to be a V phase). On the other hand, if future investigation will support the formation of a discontinuous micellar cubic phase, it might be hypothesized that, once the formation of the L$_α$ structures deprives the H phase of the more hydrophobic surfactant components, it partially converts to a concentrated, ordered micellar phase, formed by the more hydrophilic fraction of the surfactant sample. This second interpretation would explain the reason why the SLE$_1$S sample, in which more hydrophilic surfactant molecules are absent, does not form any cubic phase. In the literature, by comparing SLES samples with different average degree of ethoxylation [41] or by using computational approaches [42], it was found that, as the number of oxyethylene units increases, aggregates with higher surface curvature are favored. Interestingly, the presence of three oxyethylene units seems a ridge: SLES samples with lower ethoxylation degree form low-curvature aggregates (e.g., rods) as the concentration increases, while species with higher ethoxylation degree remain spherical [43].

Finally, independently of the structure of the cubic phase, which remains to be ascertained, molecule segregation is the only key to interpret the co-existence on three phases in a discrete range of concentration and temperature.

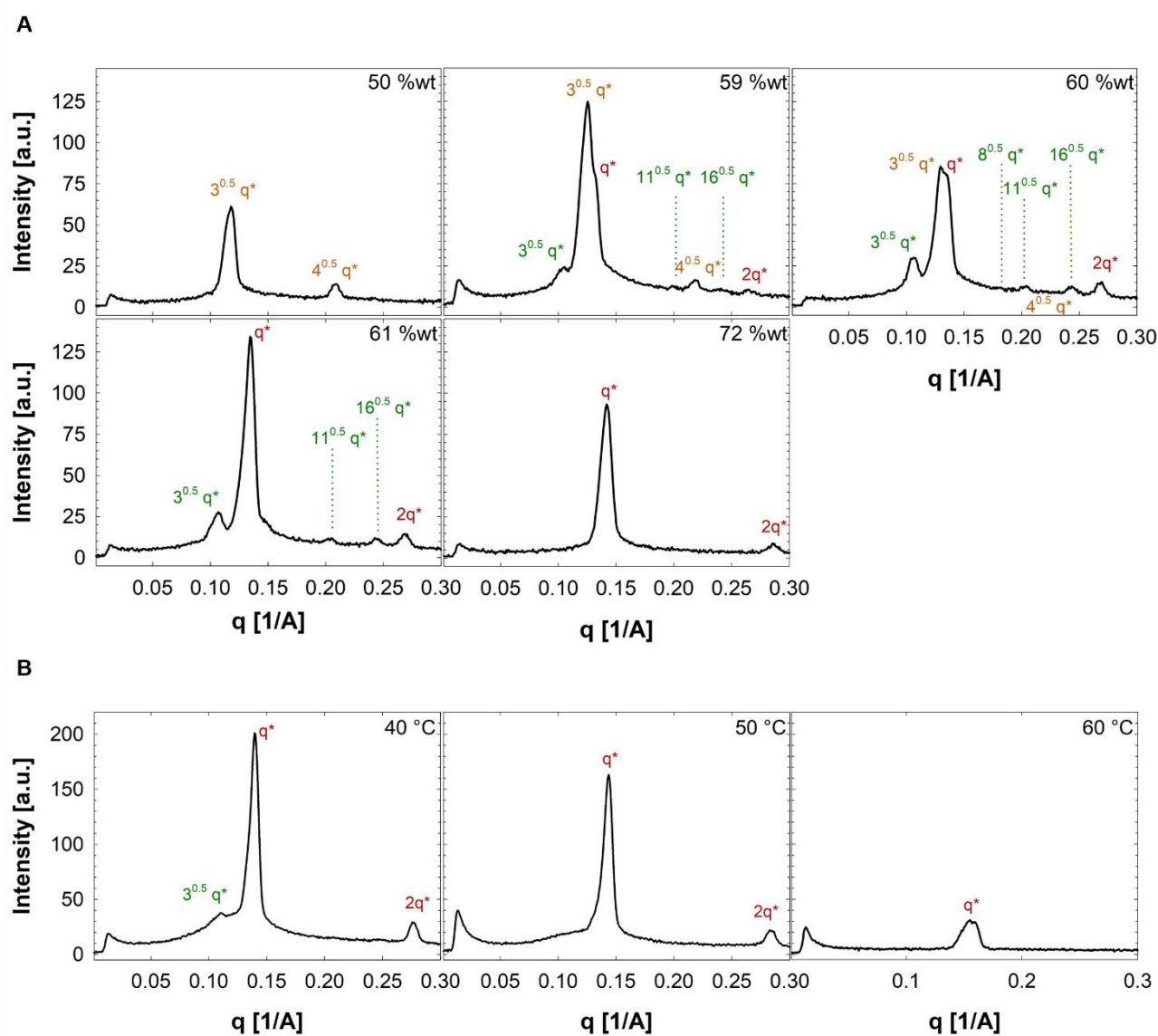

*Figure 5. SAXS profiles of SLE$_3$S-water systems. A) SAXS intensity profiles for SLE$_3$S at concentrations of 50, 59, 60, 61 and 72%wt, measured at 25 °C. The characteristic peaks corresponding to different phases are indicated using a color code: hexagonal – yellow, cubic – green), and lamellar – red. B) Temperature-dependent SAXS profiles for the SLE$_3$S at 60%wt at 40, 50, and 60°C.*

## 3.6. Rheological Perspective

To explore how the structural features of the SLE$_3$S aqueous mixtures affect the flow behavior, a rheological characterization was performed. **Figure 6** presents viscosity measurements as a function of surfactant concentration at four temperatures here in exam (30, 40, 50 and 60 °C) and at a fixed shear rate ($\dot{\gamma}$=10 1/s), in order to identify and characterize various phases.

**Figure 6A** reveals a marked increase in viscosity when transitioning from a 20%wt to a 35%wt (H) mixture across all temperatures. The viscosity remains relatively constant for the hexagonal (H) and lamellar (L$_\alpha$) phases, regardless of temperature, with a value of ~31 Pa·s at 35%wt and of ~28 Pa·s at 72%wt. However, a noticeable viscosity reduction from 15 to 8 Pa·s is observed in the 50%wt SLE$_3$S solution with increasing temperature, while the 20%wt solution shows a slight increase from 0.02 to 0.12 Pa·s, likely due to the formation of longer worm-like micelles at higher temperatures [44]. Notably, the sample at 60%wt displays a significant temperature-dependent viscosity change, dropping from 40 to 4 Pa·s between 40 and 50°C. This aligns with previous findings and is further corroborated by SAXS data, indicating that at lower temperatures, this concentration exhibits a combination of microstructures leading to higher viscosity, which gradually diminishes with increasing temperature.

A similar pattern in the relationship between viscosity and concentration is observed for both moduli, G′ and G″, as depicted in **Figure 6B** and **C**, respectively. In the 60%wt solution, both moduli radically decrease with temperature; specifically, at $\omega = 10$ rad/s, G′ falls from $3.7 \cdot 10^5$ to $6 \cdot 10^3$ Pa and G″ from $2 \cdot 10^4$ to $8.40 \cdot 10^2$ Pa. This behavior is consistent with our earlier findings and underscores the complex interplay between temperature and the microstructural composition of the surfactant solution. Conversely, for other phases (L$_1$, H and L$_\alpha$), both moduli increase by about a factor of two.

These measurements were conducted following a protocol previously established for these surfactant systems [39], where a more comprehensive rheological analysis is detailed. This ensures consistency in our methodology and allows for direct comparability with prior studies, providing a robust framework for our current findings.

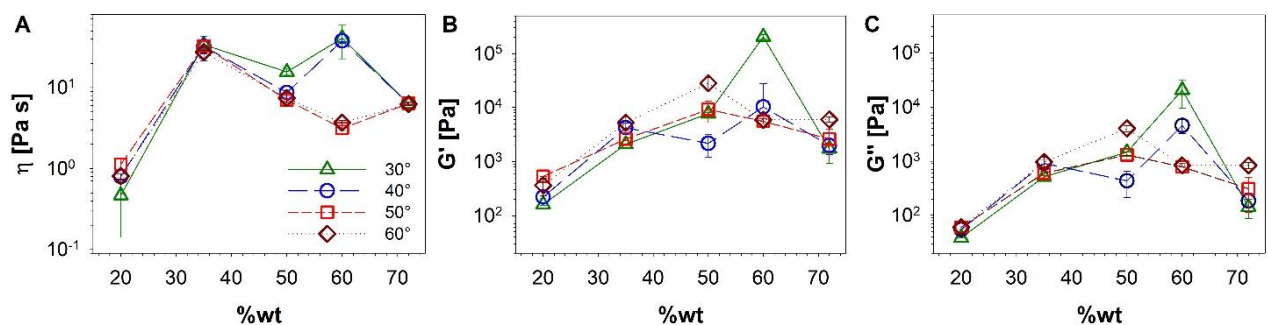

*Figure 6. Temperature and Concentration Effects on Rheological Properties of SLE$_3$S. Viscosity (η, **A**), elastic (G', **B**) and viscous (G'', **C**) moduli as a function of SLE$_3$S concentration (%wt), across different temperatures. The symbols and lines represent measurements at 30°C (triangles, solid line), 40°C (circles, dashed line), 50°C (squares, dash-dot line), and 60°C (diamonds, dotted line).*

## 3.7. Comprehensive Phase Diagram of SLE$_3$S-Water Mixtures: A Convergence of Interdisciplinary Techniques

Putting together all the information coming from the techniques used in this work, interesting details of the phase behavior of SLE$_3$S aqueous mixtures can be obtained. The SLE$_3$S-water phase diagram obtained by POM and SAXS experiments is shown in **Figure 7A**. Remarkably, a perfect agreement is found with the rheologically–based phase diagram, obtained by connecting points at the same viscosity or $G'$, as displayed in **Figure 7B**.

In samples with a surfactant concentration lower than 28%wt, isotropic mixtures are formed (L$_1$). Under crossed polarizers, these samples appear black and viscosity values are ~1 Pa·s at $\dot{\gamma}=10$ s$^{-1}$. Between 31 and 56% wt, the mixtures are birefringent and much more viscous (~20 Pa·s at $\dot{\gamma}=10$ s$^{-1}$). This phase can be identified as the H one, and this attribution is confirmed by SAXS.

Between 59 and 61% wt, in the temperature range 30-40 °C, the solutions are found to be very viscous (~40 Pa·s) and, when seen through crossed polarizers, appeared as a black background on which isolated bright domains stand out. In SAXS patterns, indeed, a coexistence between different phases – H, cubic and L$_\alpha$ – is found at room temperature, clearly distinguishable in samples at 59 and 60%wt, while H and L$_\alpha$ microstructures are identified in the 61%wt. solution. By increasing temperature above 50 °C, the cubic phase dissolved, forming an unstructured liquid, as shown by self-diffusion experiments. At the same time, sample viscosity decreased from 40 to 4 Pa·s, $G'$ decreased from 3.7·10$^5$ to 6·10$^3$ Pa and $G''$ from 2·10$^4$ to 4.5·10$^3$ Pa. These results are confirmed by visual inspection, whereby SLE$_3$S solutions at 60%wt and 60 °C are black between crossed polarizers.

In samples with a SLE$_3$S concentration higher than 62%wt, viscosity is ~6 Pa·s. These samples showed the texture of Maltese crosses characteristic of a lamellar lyotropic liquid crystalline phase. Also in this case, results are confirmed by all the techniques used in this work.

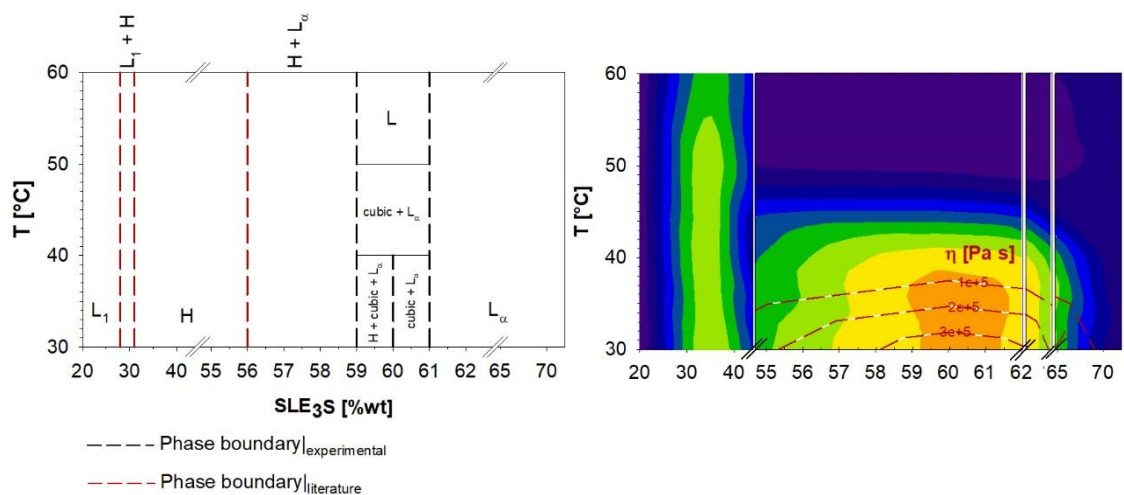

*Figure 7. Phase behavior of SLE$_3$S-water system. **A.** Compiled from experimental data, this diagram maps the phase states of the SLE$_3$S-water system across concentration and temperature ranges. Identified phases include micellar (L$_1$), hexagonal (H), cubic, and lamellar (L$_\alpha$). The 59-61%wt concentration range shows a coexistence of phases. Below 40 °C, H, cubic, and L$_\alpha$ phases are simultaneously present for 59 and 60%wt SLE$_3$S. As temperature rises, H and cubic phases transition to an isotropic liquid (L) above 40 °C and 50 °C, respectively. At 60 °C, the system transitions to the L phase. The 61%wt SLE$_3$S solution behaves similarly, with the exception that only the cubic and L$_\alpha$ phases appear below 40°C. **B.** This diagram is derived from viscosity measurements, illustrating the phase transitions within the SLE$_3$S-water system. Viscosity contours help identify the phase boundaries and transitions at different concentrations and temperatures.*

## Conclusions

A comprehensive investigation into the impact of molecular polydispersity on the phase behavior of SLES surfactants was conducted to address the knowledge gap identified in the literature. Our hypothesis is that polydispersity plays a significant role in shaping the phase behavior, with implications for both fundamental understanding and industrial applications, particularly in the formulation of detergents, personal care products, and in processes requiring specific surfactant behaviors.

Our research on SLE$_3$S allowed us to quantitatively characterize polydispersity, employing a combination of Gas Chromatography–Mass Spectrometry and Nuclear Magnetic Resonance spectroscopy. This analysis, especially in comparison with SLE$_1$S, shed light on the diverse range of ethoxylation degrees present, highlighting significant fractions of both non-ethoxylated alkyl sulfates and non-sulfonated alkylethoxylates. Thus, molecules with attitude to pack into supramolecular aggregates with different surface curvature are present.

Through Time-Lapse dissolution experiments in microchannel geometries, we mapped the phase transitions of SLE$_3$S and SLE$_1$S. A combination of techniques, including polarized light visual inspections, confocal microscopy, and Small Angle X-ray Scattering, was employed to identify specific microstructures and their concentrations. Rheological assessments across different concentrations and temperatures further elucidated the influence of these structural characteristics on flow properties.

Our findings provide a detailed analysis of the in SLE$_3$S phase transitions, with a specific focus on the concentration range of hexagonal (H), cubic, and lamellar (L$_\alpha$) stability. Notably, the coexistence of the cubic phase with both the H and L$_\alpha$ phases led to highly viscoelastic heterogeneous mixtures. The results indicated that during the transitions surfactant molecules with different packing attitude segregate, thus allowing the co-existence of aggregates with different surface curvature (e.g., cylinders, bilayers). The existence of distinct phase behavior, as revealed in polydisperse SLE$_3$S, emphasizes the profound influence of molecular polydispersity on surfactant self-assembly.

In sum, our study not only bridges a crucial knowledge gap but has also far-reaching implications for various applications [1-16]. Understanding the nuanced phase and flow behaviors of commercial surfactants is essential for optimizing product attributes and reducing environmental impacts.

## Competing interests

The authors declare no competing interests.

## Acknowledgements

We acknowledge the contribution of Pasquale Maurelli, who actively participated in the experiments and data analysis during his master's thesis. His involvement greatly contributed to the progress and outcomes of this research project.

## Author contributions

Conceptualization: GD, SC, SG; Methodology: GD, SC, SG; Formal analysis and investigation: RF, MMS, RE, SM; Writing - original draft preparation: RF; Writing - review and editing: GD, SC, SG; Supervision: SC, SG.